\input harvmac.tex
\overfullrule=0pt

\def\simge{\mathrel{%
   \rlap{\raise 0.511ex \hbox{$>$}}{\lower 0.511ex \hbox{$\sim$}}}}
\def\simle{\mathrel{
   \rlap{\raise 0.511ex \hbox{$<$}}{\lower 0.511ex \hbox{$\sim$}}}}
 
\def\slashchar#1{\setbox0=\hbox{$#1$}           
   \dimen0=\wd0                                 
   \setbox1=\hbox{/} \dimen1=\wd1               
   \ifdim\dimen0>\dimen1                        
      \rlap{\hbox to \dimen0{\hfil/\hfil}}      
      #1                                        
   \else                                        
      \rlap{\hbox to \dimen1{\hfil$#1$\hfil}}   
      /                                         
   \fi}                                         %
\def\CH{{\cal H}}

\def\CO{{\cal O}}
\def\CW{{\cal W}}
\def\ts{\thinspace}
\def\ra{\rightarrow}

\def\Lra{\Longrightarrow}

\def\leftra{\leftrightarrow}

\def\ol{\bar}

\def\gev{{\rm GeV}}
\def\tev{{\rm TeV}}

\def\half{{\textstyle{ { 1\over { 2 } }}}}
\def\third{{\textstyle{ { 1\over { 3 } }}}}
\def\fourth{{\textstyle{ { 1\over { 4 } }}}}
\def\twothirds{{\textstyle{ { 2\over { 3 } }}}}
\def\fourthirds{{\textstyle{ { 4\over { 3 } }}}}
\def\sixth{{\textstyle{ { 1\over { 6 } }}}}

\def\suc{SU(3)_C}
\def\Ntc{N}
\def\sutc{SU(N)}
\def\uone{U(1)_1}
\def\utwo{U(1)_2}
\def\uy{U(1)_Y}
\def\suone{SU(3)_1}
\def\sutwo{SU(3)_2}

\def\condab{\langle \bar T^1_L T^2_R\rangle}

\def\myfoot#1#2{{\baselineskip=14.4pt plus 0.3pt\footnote{#1}{#2}}}

\Title{\vbox{\baselineskip12pt\hbox{BUHEP--98--9}
\hbox{hep-ph/9805254}}}
\centerline{\titlefont{A New Model of Topcolor-Assisted Technicolor}}

\bigskip
\centerline{Kenneth Lane\myfoot{$^{\dag }$}{lane@buphyc.bu.edu}}
\smallskip\centerline{Department of Physics, Boston University}
\centerline{590 Commonwealth Avenue, Boston, MA 02215}
\vskip .3in

\centerline{\bf Abstract}

I present a model of topcolor-assisted technicolor that can have topcolor
breaking of the desired pattern, hard masses for all quarks and leptons,
mixing among the heavy and light generations, and explicit breaking of all
technifermion chiral symmetries except electroweak $SU(2) \otimes $U(1).
These positive features depend on the outcome of vacuum alignment. The main
flaw in this model is tau-lepton condensation.

\bigskip


\vfil\eject

It is not difficult to construct a dynamical model of electroweak 
symmetry breaking. We have known how at least since 1973
\ref\tc{M.~Weinstein, Phys.~Rev.~{\bf D8}, 2511 (1973)\semi
S.~Weinberg, Phys.~Rev.~{\bf D19}, 1277 (1979)\semi
L.~Susskind, Phys.~Rev.~{\bf D20}, 2619 (1979).}.
The difficulties lie in extending this dynamics to flavor: accounting for
the masses of all known fermions, including the top quark's; breaking
technifermion chiral symmetries to prevent light technipions with
axion-strength couplings to quarks and leptons; and evading the many
phenomenological pitfalls---flavor-changing neutral currents to name the
most famous and ubiquitous example---that plague any theory of flavor
\ref\etc{E.~Eichten and K.~Lane, Phys.~Lett.~{\bf B90}, 125 (1980).}.
This paper develops further the topcolor-assisted technicolor approach to
accomplishing all this.

Topcolor-assisted technicolor (TC2) is the only scheme known in which 
there is an explicit dynamical and natural mechanism for breaking 
electroweak symmetry and generating the fermion masses including
$m_t \simeq 175\,\gev$. In TC2, there are no elementary scalar fields
and no unnatural or excessive fine-tuning of parameters
\ref\tctwohill{C.~T.~Hill, hep-ph/9411426, Phys.~Lett.~{\bf B345}, 483
(1995).}.
In Hill's simplest TC2 model, the third generation of quarks and leptons 
transforms under strongly-coupled color and hypercharge groups, $\suone
\otimes \uone$, with the usual charges, while the light generations
transform under weakly-coupled $\sutwo \otimes \utwo$. Near $1\,\tev$,
these four groups are broken, somehow, to the diagonal subgroup of ordinary
color and hypercharge, $\suc\otimes \uy$. The desired pattern of heavy
quark condensation occurs because $\uone$ couplings are such that the
spontaneously broken $\suone \otimes \uone$ interactions are supercritical
only for the top quark.

Hill did not address the issues of topcolor breaking, generational mixing
and chiral symmetry breaking. In addition to these concerns, there are
stringent constraints on model-building from the conflict between custodial
isospin conservation and the large topcolor $\uone$ coupling
\ref\cdt{R.~S.~Chivukula, B.~A.~Dobrescu and J.~Terning, hep-ph/9503203,
Phys.~Lett.~{\bf B353}, 289 (1995).},
and from limits on $B_d-\ol B_d$ mixing
\ref\kominis{D.~Kominis, hep-ph/9506305, Phys.~Lett.~{\bf B358}, 312
(1995)\semi G.~Buchalla, G.~Burdman, C.~T.~Hill and D.~Kominis,
hep-ph/9510376, Phys.~Rev.~{\bf D53}, 5185 (1996).}.
These constraints, the cancellation of $U(1)$ anomalies, and the dynamics
of generational mixing and topcolor breaking to $\suc \otimes \uy$ were
considered in Refs.~\ref\tctwoklee{K.~Lane and E.~Eichten, hep-ph/9503433,
Phys.~Lett.~{\bf B352}, 382 (1995).} and~\ref\tctwokl{K.~Lane,
Phys.~Rev.~{\bf D54}, 2204 (1996); also {\it Progress on Symmetry Breaking
and Generational Mixing in Topcolor-Assisted Technicolor}, hep-ph/9703233,
in the Proceedings of the 1996 Workshop on Strongly Coupled Gauge
Theories, Nagoya, Japan, (November 1996).}. The main features of the models
developed in these studies are:

\item{(1)} The $\uone$ charges of technifermions are custodial-isospin
symmetric.

\item{(2)} Above the electroweak scale, third-generation quarks transform
under strongly-coupled $\suone$ while the two light-generation quarks
transform under the weaker $\sutwo$. However, {\it all} quarks and leptons
transform under the strongly-coupled $\uone$.

\item{(3)} In order that $Z^0$ couplings be nearly standard, the breakdown
$\uone \otimes \utwo \ra \uy$ necessarily occurs at a somewhat higher scale
than $SU(2) \otimes \uy \ra U(1)_{EM}$. This is effected by a higher
dimensional technifermion $\psi$ whose condensate is $SU(2) \otimes \uy$
invariant. The $\psi$-condensate gives rise to a 2--3~TeV $Z'$ boson with
much interesting phenomenology
\cdt, \kominis,
\ref\sbl{Y.~Su, G.~F.~Bonini and K.~Lane, hep-ph/9706267,
Phys.~Rev.~Lett.~{\bf 79}, 4075 (1997).},
\ref\rador{T.~Rador, {\it Rare Process Constraints on the Topcolor $Z'$
Boson}, in preparation.},
\ref\cps{M.~Popovic, E.~H.~Simmons and R.~S.~Chivukula, {\it A Heavy Top
Quark from Flavor-Universal Colorons}, BUHEP--98--10, in preparation.}.

\item{(4)} The breaking of the color and electroweak symmetries to $\suc
\otimes U(1)_{EM}$ is due to technifermions in the fundamental
representation of the TC gauge group, assumed to be $SU(N)$. In particular,
the $\suone \otimes \sutwo$ breaking condensate $\condab$, where $T^i$ is a
triplet of $SU(3)_i$, is driven by an attractive strong $\uone$
interaction.

\item{(5)} Generational mixing is produced by an extended technicolor (ETC)
operator which induces the transition $d_L,s_L \leftra b_R$, but {\it not}
$d_R,s_R \leftra b_L$. In this way, the excessive $B_d-\ol B_d$ mixing
discussed in Ref.~\kominis\ is avoided.

\item{(6)} Nontrivial solutions exist to all the $U(1)$
anomaly-cancellation equations.

\noindent These constraints led to a proliferation of technifermions and a
large chiral $SU(N_T)_L \otimes SU(N_T)_R$ symmetry. In the models
considered, it was not possible to break explicitly all unwanted chiral
symmetries, so that massless or very light technipions occurred.

Explicit chiral symmetry breaking and generational mixing, in the form of
quark mass $\ol q T \ol T q$ and technipion mass $\ol T T \ol T T$
operators, are induced mainly by ETC interactions. Here, $T = (U,D)$ are
technifermion isodoublets. Let us define a ``complete set'' of $SU(2)
\otimes U(1)$-invariant 4T operators $\ol T^i_L \gamma^\mu T^j_L \ts\ts \ol
T^k_R \gamma_\mu (a + b \sigma_3) T^l_R$ as one for which no technifermion
global symmetry generator commutes with every member of the set. In a
complete set, every left-handed and right-handed technifermion field
appears in at least one of the operators. (This excludes operators in which
the left or right-handed currents involve the same technifermion twice,
e.g., operators generated by diagonal ETC or $\uone$ interactions.) Since I
have not specified an ETC group and its breaking pattern, it is necessary
to assume that the required operators exist, {\it provided} they respect
all known gauge interactions, including $\uone \otimes \utwo$. For the type
of model considered in Ref.~\tctwokl, I was unable to find a complete set
of 4T operators.

Even a complete set of operators is not sufficient to guarantee
that all technipion masses are large. It is also necessary that condensates
form so that all 4T operators have nonzero vacuum expectation values, i.e.,
that they contribute to the vacuum energy
\eqn\vac{E(\CW) = \langle \Omega \vert \CW \CH' \CW^{-1} \vert \Omega
\rangle \ts.}
Here, the hamiltonian $\CH'$ is the sum over all allowed 4T operators and
$\CW$ is an $SU(N_T)_L \otimes SU(N_T)_R$ transformation. Finding the
transformation $\CW^0$ which minimizes $E(\CW)$ is known as vacuum
alignment
\ref\dashen{R.~Dashen Phys.~Rev.~{\bf D3}, 1879 (1971).}.
In the correct vacuum, $\langle \ol T^i_L T^j_R \rangle \propto
W^0_{ij}$, where $W^0$ is the corresponding $SU(N_T)$ matrix. The
models under consideration have a large number of technifermions and 4T
operators, and minimization is a complicated numerical task, now under
study.

I present here a type of TC2 model which does allow a complete set of 4T
operators.  For the models of Ref.~\tctwokl, the difficulty of constructing
such a set was due at least in part to the fact that light and heavy quarks
transform under different color $SU(3)$ groups. Then their hypercharges
were tightly constrained by cancellation of $U(1) \ts [SU(3)]^2$ anomalies
and there was no complete set invariant under $\uone \otimes \utwo$. In the
model presented here, I adopt the ``flavor-universal topcolor'' of
Chivukula, Cohen and Simmons~\ref\ccs{R.~S.~Chivukula, A.~G.~Cohen and
E.~H.~Simmons, Phys.~Lett.~{\bf B380}, 92 (1996).}; also see Ref.~\cps.
Their model was motivated by the apparent excess of high-$E_T$ events in
the CDF jet data~\ref\CDF{F.~Abe, et al., CDF Collaboration,
hep-ex/9601008, Phys.~Rev.~Lett.~{\bf 77}, 438 (1996).}. They used two
$SU(3)$ groups, but assumed all quarks transform under only the stronger
$\suone$ color group. I find that this allows simpler quark hypercharges
than in Ref.~\tctwokl\ and, so, the $U(1)$ constraints for a complete set
of 4T operators can be met. A dynamical mechanism for breaking $\suone
\otimes \sutwo \ra \suc$ was not provided in Refs.~\cps\ and \ccs. I shall
use the condensation of technifermions transforming under the two color
groups to effect this breaking. The model I present has one obvious bad
feature: the tau-lepton has very strong, attractive $\uone$ interactions
and, therefore, it has a large condensate and mass.~\foot{As I once heard
in a similar situation, ``the tau-lepton is the {\it bane in mayn haldz}''
(the bone in my throat). At least, the Goldstone boson from
tau-condensation acquires a sizable mass from the ETC contribution to
$m_\tau$.}

The fermions in this new model, their color representations and $U(1)$
charges are listed in Table~1. Technifermions $T^i_{L,R}$ transform under
$SU(N)$ as fundamentals, while $\psi_{L,R}$ are antisymmetric tensors. As
noted above, the condensate $\langle \ol  \psi_L \psi_R \rangle$ breaks
$\uone \otimes \utwo \ra \uy$ and $\condab  \neq 0$ breaks $\suone \otimes
\sutwo \ra \suc$. The condition on the $\uone$ charges of $T^1$ and $T^2$
required to form this condensate is, in the walking technicolor and
large-$N$ limits~\tctwokl,
\eqn\subreak{(u_1 - v_1) (v_1 - u'_1) > {4\alpha_3 \over {3\alpha_1}} \ts.}
Here, $\alpha_3$ and $\alpha_1$ are the $\suone$ and $\uone$ couplings 
near 1~TeV. Note that this requires and $(u_1 - u'_1)^2 > (u_1 - v_1) (v_1
- u'_1) > 0$. To preserve $U(1)_{EM}$, $T^1$ and $T^2$ must have equal
electric charges, i.e., $u_1 + u_2 = u'_1 + u'_2 = v_1 + v_2$

To give mass to quarks and leptons, I assume the following ETC operators:
\eqn\qTTq{\eqalign{
\ol \ell^l_{iL} \gamma^\mu T^l_L \ts\ts \ol D^l_R \gamma_\mu e_{jR}
\qquad &\Lra \qquad x_1 - x'_1 =0 \cr
\ol q^l_{iL} \gamma^\mu T^l_L \ts\ts \ol T^l_R \gamma_\mu q^l_{jR} 
\qquad &\Lra \qquad x_1 - x'_1 = 0 \cr
\ol \ell^h_L \gamma^\mu T^t_L \ts\ts \ol D^t_R \gamma_\mu \tau_R
\qquad &\Lra \qquad a - a' = y_1 - y'_1 \cr
\ol q^h_L \gamma^\mu T^t_L \ts\ts \ol U^t_R \gamma_\mu t_R
\qquad &\Lra \qquad b - b' = y_1 - y'_1 \cr
\ol q^h_L \gamma^\mu T^b_L \ts\ts \ol D^b_R \gamma_\mu b_R
\qquad &\Lra\qquad b - b'' = z_1 - z'_1 \ts\ts. \cr}}
To generate $d_L,s_L \leftra b_R$, I require the operator~\foot{This choice
is not unique. Two other operators are consistent with the hypercharge
conditions implied by the ETC operators and the anomaly constraints. They
are $\ol q^l_{iL} \gamma^\mu T^l_L \ts\ts \ol D^t_R \gamma_\mu b_R$ and
$\ol q^l_{iL} \gamma^\mu T^t_L \ts\ts \ol D^t_R \gamma_\mu b_R$. These
generation-mixing operators are not simultaneously consistent with
the $U(1)$ symmetries; only one may be assumed to exist.}
\eqn\sLbR{\ol q^l_{iL} \gamma^\mu T^t_L \ts\ts \ol D^b_R \gamma_\mu b_R
\qquad \Lra \qquad b'' = z'_1 - y_1 \ts\ts.}
Of course, the technifermions in these operators must condense in the
correctly aligned ground state. To forbid the transition $d_R,s_R \leftra
b_L$ and unacceptably large $B_d - \ol B_d$ mixing, ETC interactions must
not generate any of the operators $\ol q^h_L \gamma^\mu T^i_L \ts\ts \ol
D^j_R \gamma_\mu d^l_{R}$ for any $i,j$. This gives the constraints $b \neq
0, u_1 - u'_1, y_1 - y'_1, z_1 - z'_1$, etc.

A complete set of allowed $SU(2) \otimes U(1)$-invariant 4T operators is
\eqn\TTTT{\eqalign{
\ol T^1_L \gamma^\mu T^l_L \ts\ts \ol T^b_R \gamma_\mu T^1_R \qquad
&\Lra \qquad u_1 - u'_1 = x_1 - z'_1\cr
\ol T^1_L \gamma^\mu T^b_L \ts\ts \ol T^t_R \gamma_\mu T^1_R \qquad
&\Lra \qquad u_1 - u'_1 = z_1 - y'_1\cr
\ol T^t_L \gamma^\mu T^b_L \ts\ts \ol T^b_R \gamma_\mu T^l_R \qquad
&\Lra \qquad z_1 - z'_1  = y_1 - x'_1 = 0 \cr
\ol T^l_L \gamma^\mu T^b_L \ts\ts \ol T^t_R \gamma_\mu T^b_R \qquad
&\Lra \qquad x_1 - z'_1  = z_1 - y'_1 \cr
\ol T^2_L \gamma^\mu T^l_L \ts\ts \ol T^l_R \gamma_\mu T^2_R \qquad
&\Lra \qquad x_1 - x'_1 = 0 \ts\ts. \cr}}
Note that the equal-charge conditions $x_1 + x_2 = y_1 +  y_2 = z_1 + z_2$
are implied by these operators. In addition to this set, diagonal 4T
operators from broken ETC and $\uone$ interactions contribute to the
chiral-breaking hamiltonian, $\CH'$.

The requirement that gauge anomalies cancel further constrains $U(1)$
charge assignments. Taking account of the equal-charge conditions, there
are four independent conditions which are linear in the hypercharges (the
$U(1)_{1,2}[\sutwo]^2$ condition is automatically satisfied):
\eqn\linanom{\eqalign{
&{\underline{U(1)_{1,2} [\sutc]^2:}}
\qquad 3(u_1 - u'_1) + y_1 - y'_1 + z_1 - z'_1
= -\half (\Ntc -2) (\xi - \xi') \cr\cr
&{\underline{U(1)_{1,2} [\suone]^2:}}
\qquad 2b - b' - b'' = -2\Ntc(u_1-u'_1) \cr\cr
&{\underline{U(1)_{1,2} [SU(2)]^2:}} 
\qquad \hskip 0.2truein a+3b = -\Ntc [3(u_1 + v_1) + x_1 + y_1 + z_1]
\cr
&\hskip2.2truein =
\Ntc [3(u_2 + v_2) + x_2 + y_2 + z_2] \ts\ts. \cr}}
Taken together with the hypercharge conditions, Eqs.~\qTTq\ and \TTTT, 
there follow the relations:
\eqn\relations{\eqalign{
& b = -(u_1 - u'_1) \ts, \quad b' = -3(u_1 - u'_1) \ts, \quad
b'' = z'_1 - y_1 = (2N+1) (u_1 - u'_1) \cr
& a - a' = b - b' =  y_1 - y'_1  = 2(u_1 - u'_1) \cr
& b - b'' = z_1 - z'_1 = -2(N+1)(u_1 - u'_1) \cr
& \xi - \xi' = 2\left({2N-3\over{N-2}}\right) (u_1 - u'_1) \ts.\cr}}
Note that $bb' > 0$ and $bb'' < 0$ which favors top, but not bottom,
condensation.

There are four anomaly conditions that are cubic in the
hypercharges. However, the $\uy[SU(2)]^2$ anomaly cancellation
guarantees that the $[\uy]^3$ anomaly also cancels, leaving three
independent conditions. They are conveniently given for $[\uone]^3$,
$[\uone]^2 \uy$, and $[\uone]^3 +  [\utwo]^3 - 3[\uone]^2 \uy$:
\eqn\cubanom{\eqalign{
& \quad 0 = 2\Ntc\biggl[3(u^3_1 - u^{\prime 3}_1)
+ y^3_1 - y^{\prime 3}_1  + z^3_1 - z^{\prime 3}_1\biggr] \cr
& \qquad + \half \Ntc(\Ntc - 1)(\xi^3 - \xi^{\prime 3}) +
2a^3 - a^{\prime 3} +3(2 b^3 - b^{\prime 3} - b^{\prime\prime 3})\cr\cr
&\quad 0 = 2\Ntc \biggl[3(u_1 + u_2) (u^2_1 - u^{\prime 2}_1)
+ (y_1 + y_2) (y^2_1 - y^{\prime 2}_1)
+ (z_1 + z_2) (z^2_1 - z^{\prime 2}_1)\biggr] \cr
& \qquad + a^{\prime 2} - a^2 + b^2 - 2 b^{\prime 2} +
b^{\prime\prime 2} \cr\cr
& \quad 0 =  2\Ntc\biggl\{3(u'_1 - u_1) \ts\left[(u_1+u_2)^2
+\fourth\right]
+ (y'_1 - y_1) \ts\left[(y_1+y_2)^2 +\fourth\right] \cr
& \qquad \qquad \ts\ts\ts
+ (z'_1 - z_1) \ts\left[(z_1+z_2)^2 +\fourth\right]\biggr\}
+ a - a' + \fourthirds (b - b') + \third (b - b'') \ts\ts. \cr}}

These conditions have an infinite number of solutions. Following
Ref.~\tctwokl, I found one with $\vert u_1 - u'_1\vert = \CO(1)$, as
required for naturally large couplings in Eq.~\subreak, as follows: I
assumed $u_1 = - u'_1$ and $\xi = - \xi'$. Then, for $N=4$, I chose $z_1
=8$ and $z_1 + z_2 = 2$. This input has the nontrivial solution
\eqn\soln{a = - 15.437 \ts, \qquad u_1 = - u'_1 = -0.648 \ts, \qquad u_1 +
u_2 = 3.321 \ts.}
The other hypercharges are to be chosen in accord with
Eqs.\qTTq--\relations.

The large values $a \simeq a' \simeq -2 z_1$ found in the solutions to
Eqs.~\cubanom\ are unavoidable: The $[\uone]^3$ condition has no real
solution for $u_1 - u'_1 \neq 0$ and $\vert a \vert \simle \vert u_1 - u'_1
\vert$. The large positive value of $aa'$ then suggests that the $\uone$
interactions generate a tau-condensate $\langle \ol \tau_L \tau_R \rangle
\sim \langle \ol t_L t_R \rangle$. Such a hypercharge also raises the
question of the triviality of the $\uone$ interaction: does the Landau pole
occur at an energy significantly lower than the one at which we can
envisage $\uone$ being unified into an asymptotically free ETC
group~\tctwokl? I know of no choice of chiral symmetry breaking ETC
operators and associated hypercharge assignments within the present simple
model of flavor-universal topcolor which evades $aa'/(u_1 - u'_1)^2 \gg 1$.
It may be possible to find an acceptable model, including a complete set of
4T operators, by enlarging the technifermion sector and/or complicating the
light generation hypercharge assignments.

In conclusion, I have constructed a TC2 model with flavor-universal
topcolor that seems capable of satisfying all major phenomenological
constraints except those involving the tau-lepton. To my mind, the more
important task ahead is to show that a nontrivial vacuum-alignment solution
exists that results in nonzero masses and mixings for all the fundamental
fermions and composite technipions.

I am grateful to E.~Simmons for a careful reading of the manuscript and
valuable comments. This research was supported in part by the Department of
Energy under Grant~No.~DE--FG02--91ER40676.

\listrefs

\centerline{\vbox{\offinterlineskip
\hrule\hrule\hrule
\halign{&\vrule#&
  \strut\quad#\hfil\quad\cr
height4pt&\omit&&\omit&&\omit&&\omit&&\omit&\cr\cr
&\hfill Particle \hfill&&\hfill$SU(3)_1$ \hfill&&\hfill
 $SU(3)_2$\hfill&&\hfill$Y_1$\hfill&&\hfill $Y_2$\hfill &\cr\cr
height4pt&\omit&&\omit&&\omit&&\omit&&\omit&\cr\cr
\noalign{\hrule\hrule\hrule}
height4pt&\omit&&\omit&&\omit&&\omit&&\omit&\cr\cr
&$\ell_L^l$&&\hfill$1$\hfill&&\hfill$1$\hfill
&&\hfill$0$\hfill&&\hfill$-\half$\hfill&\cr\cr
\noalign{}
height4pt&\omit&&\omit&&\omit&&\omit&&\omit&\cr\cr
&$e_R$, $\mu_R$&&\hfill$1$\hfill&&\hfill$1$\hfill
&&\hfill$0$\hfill&&\hfill$-1$\hfill&\cr\cr
\noalign{\hrule}
height4pt&\omit&&\omit&&\omit&&\omit&&\omit&\cr\cr
&$q_L^l$&&\hfill$3$\hfill&&\hfill$1$\hfill
&&\hfill$0$\hfill&&\hfill$\sixth$\hfill&\cr\cr
\noalign{}
height4pt&\omit&&\omit&&\omit&&\omit&&\omit&\cr\cr
&$u_R$, $c_R$&&\hfill$3$\hfill&&\hfill$1$\hfill
&&\hfill$0$\hfill&&\hfill$\twothirds$\hfill&\cr\cr
\noalign{}
height4pt&\omit&&\omit&&\omit&&\omit&&\omit&\cr\cr
&$d_R$, $s_R$&&\hfill$3$\hfill&&\hfill$1$\hfill
&&\hfill$0$\hfill&&\hfill$-\third $\hfill&\cr\cr
\noalign{\hrule}
height4pt&\omit&&\omit&&\omit&&\omit&&\omit&\cr\cr
&$\ell_L^h$&&\hfill$1$\hfill&&\hfill$1$\hfill
&&\hfill$a$\hfill&&\hfill$-\half-a$\hfill&\cr\cr
\noalign{}
height4pt&\omit&&\omit&&\omit&&\omit&&\omit&\cr\cr
&$\tau_R$&&\hfill$1$\hfill&&\hfill$1$\hfill
&&\hfill$a'$\hfill&&\hfill$-1-a'$\hfill&\cr\cr
\noalign{\hrule}
height4pt&\omit&&\omit&&\omit&&\omit&&\omit&\cr\cr
&$q_L^h$&&\hfill$3$\hfill&&\hfill$1$\hfill
&&\hfill$b$\hfill&&\hfill$\sixth-b$\hfill&\cr\cr
\noalign{}
height4pt&\omit&&\omit&&\omit&&\omit&&\omit&\cr\cr
&$t_R$&&\hfill$3$\hfill&&\hfill$1$\hfill
&&\hfill$b'$\hfill&&\hfill$\twothirds-b'$\hfill&\cr\cr
\noalign{}
height4pt&\omit&&\omit&&\omit&&\omit&&\omit&\cr\cr
&$b_R$&&\hfill$3$\hfill&&\hfill$1$\hfill
&&\hfill$b''$\hfill&&\hfill$-\third - b''$\hfill&\cr\cr
\noalign{\hrule\hrule}
height4pt&\omit&&\omit&&\omit&&\omit&&\omit&\cr\cr
&$T_L^1$&&\hfill$3$\hfill&&\hfill$1$\hfill
&&\hfill$u_1$\hfill&&\hfill$u_2$\hfill&\cr\cr
\noalign{}
height4pt&\omit&&\omit&&\omit&&\omit&&\omit&\cr\cr
&$U_R^1$&&\hfill$3$\hfill&&\hfill$1$\hfill
&&\hfill$u'_1$\hfill&&\hfill$u'_2+\half$\hfill&\cr\cr
\noalign{}
height4pt&\omit&&\omit&&\omit&&\omit&&\omit&\cr\cr
&$D_R^1$&&\hfill$3$\hfill&&\hfill$1$\hfill
&&\hfill$u'_1$\hfill&&\hfill$u'_2-\half$\hfill&\cr\cr
\noalign{\hrule}
height4pt&\omit&&\omit&&\omit&&\omit&&\omit&\cr\cr
&$T_L^2$&&\hfill$1$\hfill&&\hfill$3$\hfill
&&\hfill$v_1$\hfill&&\hfill$v_2$\hfill&\cr\cr
\noalign{}
height4pt&\omit&&\omit&&\omit&&\omit&&\omit&\cr\cr
&$U_R^2$&&\hfill$1$\hfill&&\hfill$3$\hfill
&&\hfill$v_1$\hfill&&\hfill$v_2+\half$\hfill&\cr\cr
\noalign{}
height4pt&\omit&&\omit&&\omit&&\omit&&\omit&\cr\cr
&$D_R^2$&&\hfill$1$\hfill&&\hfill$3$\hfill
&&\hfill$v_1$\hfill&&\hfill$v_2-\half$\hfill&\cr\cr
\noalign{\hrule\hrule}
height4pt&\omit&&\omit&&\omit&&\omit&&\omit&\cr\cr
&$T_L^l$&&\hfill$1$\hfill&&\hfill$1$\hfill
&&\hfill$x_1$\hfill&&\hfill$x_2$\hfill&\cr\cr
\noalign{}
height4pt&\omit&&\omit&&\omit&&\omit&&\omit&\cr\cr
&$U_R^l$&&\hfill$1$\hfill&&\hfill$1$\hfill
&&\hfill$x'_1$\hfill&&\hfill$x'_2+\half$\hfill&\cr\cr
\noalign{}
height4pt&\omit&&\omit&&\omit&&\omit&&\omit&\cr\cr
&$D_R^l$&&\hfill$1$\hfill&&\hfill$1$\hfill
&&\hfill$x'_1$\hfill&&\hfill$x'_2-\half$\hfill&\cr\cr
\noalign{\hrule}
height4pt&\omit&&\omit&&\omit&&\omit&&\omit&\cr\cr
&$T_L^t$&&\hfill$1$\hfill&&\hfill$1$\hfill
&&\hfill$y_1$\hfill&&\hfill$y_2$\hfill&\cr\cr
\noalign{}
height4pt&\omit&&\omit&&\omit&&\omit&&\omit&\cr\cr
&$U_R^t$&&\hfill$1$\hfill&&\hfill$1$\hfill
&&\hfill$y'_1$\hfill&&\hfill$y'_2+\half$\hfill&\cr\cr
\noalign{}
height4pt&\omit&&\omit&&\omit&&\omit&&\omit&\cr\cr
&$D_R^t$&&\hfill$1$\hfill&&\hfill$1$\hfill
&&\hfill$y'_1$\hfill&&\hfill$y'_2-\half$\hfill&\cr\cr
\noalign{\hrule}
height4pt&\omit&&\omit&&\omit&&\omit&&\omit&\cr\cr
&$T_L^b$&&\hfill$1$\hfill&&\hfill$1$\hfill
&&\hfill$z_1$\hfill&&\hfill$z_2$\hfill&\cr\cr
\noalign{}
height4pt&\omit&&\omit&&\omit&&\omit&&\omit&\cr\cr
&$U_R^b$&&\hfill$1$\hfill&&\hfill$1$\hfill
&&\hfill$z'_1$\hfill&&\hfill$z'_2+\half$\hfill&\cr\cr
\noalign{}
height4pt&\omit&&\omit&&\omit&&\omit&&\omit&\cr\cr
&$D_R^b$&&\hfill$1$\hfill&&\hfill$1$\hfill
&&\hfill$z'_1$\hfill&&\hfill$z'_2-\half$\hfill&\cr\cr
\noalign{\hrule\hrule}
height4pt&\omit&&\omit&&\omit&&\omit&&\omit&\cr\cr
&$\psi_L$&&\hfill$1$\hfill&&\hfill$1$\hfill
&&\hfill$\xi$\hfill&&\hfill$-\xi$\hfill&\cr\cr
\noalign{}
height4pt&\omit&&\omit&&\omit&&\omit&&\omit&\cr\cr
&$\psi_R$&&\hfill$1$\hfill&&\hfill$1$\hfill
&&\hfill$\xi'$\hfill&&\hfill$-\xi'$\hfill&\cr\cr
height4pt&\omit&&\omit&&\omit&&\omit&&\omit&\cr\cr}
\hrule\hrule\hrule}}

\centerline{TABLE 1: Lepton, quark and technifermion colors and
hypercharges.}

\vfil\eject

\bye